\documentclass[journal,a4paper]{IEEEtran}%

\usepackage{graphicx}
\usepackage{amsmath}
\usepackage{amsfonts}
\usepackage{amssymb}
\usepackage[utf8]{inputenc}
\usepackage{cite}
\usepackage{ntheorem}
\usepackage{algorithm}
\usepackage{algorithmicx}
\usepackage{algpseudocode}
\usepackage{listings}
\usepackage{cite}
\usepackage{xcolor}
\usepackage{url}
\usepackage{ntheorem}
\usepackage{algorithm}
\usepackage{algorithmicx}
\usepackage{algpseudocode}

\usepackage{tabularx}
\usepackage{booktabs,makecell}

\IEEEoverridecommandlockouts

\algnewcommand\algorithmicoutput{\textbf{Output:}} 
\algnewcommand\Output{\item[\algorithmicoutput]}
\algnewcommand\algorithmicinput{\textbf{Input:}} 
\algnewcommand\Input{\item[\algorithmicinput]}

\newcounter{theRemark}
\newenvironment{Remark}{
	\par\smallskip\refstepcounter{theRemark}%
	\noindent%
	\textbf{Remark~\arabic{theRemark}}:~%
	\ignorespaces%
}{
	\par\smallskip
}

\begin{document}

\title{Quantization in Compressive Sensing: \\ A Signal Processing Approach}
\author{
Isidora Stankovi\'c,~\IEEEmembership{Student Member,~IEEE,} 
Milo\v s Brajovi\'c,~\IEEEmembership{Member,~IEEE,} \\
Milo\v s Dakovi\'c,~\IEEEmembership{Member,~IEEE,}
Cornel Ioana,~\IEEEmembership{Member,~IEEE,}
Ljubi\v{s}a Stankovi\'c,~\IEEEmembership{Fellow,~IEEE}
\thanks{
\ I. Stankovi\'c, M. Brajovi\'c, M. Dakovi\'c, L. Stankovi\'c are with the Faculty of Electrical Engineering, University of Montenegro, D\v{z}ord\v{z}a Va\v{s}ingtona bb, 81000 Podgorica, Montenegro, e-mail: \{isidoras, milos, milosb, ljubisa\}@ucg.ac.me. 

I. Stankovi\'c and C. Ioana, are with GIPSA Lab, INP Grenoble, University of Grenoble Alpes, 11 Rue des Math\'ematiques, 38000 Grenoble, France, e-mail: cornel.ioana@gipsa-lab.grenoble-inp.fr.}}

\maketitle

\begin{abstract}
Influence of the finite-length registers and quantization effects on the reconstruction of sparse and approximately sparse signals is analyzed in this paper. For the nonquantized measurements, the  compressive sensing (CS) framework provides highly accurate reconstruction algorithms that produce negligible errors when the reconstruction conditions are met. However, hardware implementations of signal processing algorithms involve the finite-length registers and quantization of the measurements. An analysis of the effects related to the measurements quantization with an arbitrary number of bits is the topic of this paper. A unified mathematical model for the analysis of the quantization noise and the signal nonsparsity on the CS reconstruction is presented.  An exact formula for the expected energy of error in the CS-based reconstructed signal is derived. The theory is validated through various numerical examples with quantized measurements, including the cases of approximately sparse signals, noise folding, and floating-point arithmetics.

\end{abstract}

\section{Introduction}
\label{Intro}
\IEEEPARstart{C}{ompressive} sensing (CS) theory provides a rigorous mathematical framework for the reconstruction of sparse signals, using a reduced set of measurements  \cite{Donoho_CS,Baraniuk_CS, davenport, L1_magic,candesQUANT, candesQUANT2, Srdjan_CS,Uniq,IETsigProc}. Advantages of CS are directly related to the signal transmission and storage efficiency, which is crucial in big data setups. Moreover, the problem of the physical unavailability of measurements, or the problem of a significant signal corruption, are also potentially solvable within the CS framework. Since the establishment of CS, the phenomena related to the reduced sets of measurements and sparse signal reconstruction have been supported by the fundamental theory and well-defined mathematical framework,
 while the performance of the reconstruction processes have been continuously improved by newly introduced algorithms, often adapted to perform in a particular context, or to solve specific problems \cite{cosamp,D1,IHTpaper,geosc,DSPknjiga,tutorial}. In real applications, many signals are sparse or approximately sparse in a certain transformation domain. This makes the CS applicable in various fields of signal processing \cite{tutorial}. 

Ideally, the measurements used for the reconstruction should be taken accurately, assuming a very large number of bits in their digital form (providing high precision levels). However, this could be extremely demanding and expensive for  hardware implementations \cite{1bitCS}. Therefore, in practice, the measurements are quantized, meaning that they are represented using a limited number of bits. Such measurements bring robustness, memory efficiency and simplicity in the corresponding hardware  implementation (particularly in sensor design). This paper investigates the quantization influence on the CS reconstruction with a simple yet rigorous characterization of the related phenomena, through the derivation of the corresponding errors. The theory is supported by a relevant theoretical framework and detailed statistical analysis, through  extensive numerical experiments. 

The most extreme case of quantization is limiting the measurements using only one bit. In previous work \cite{1bitCS,BIHTpaper,NorthThesis,Greedy1bit}, one-bit measurements are initially treated as sign constraints, as opposed to the values to be matched in the mean squared sense during the reconstruction process.  Quantization to one-bit measurements is suitable for hardware systems since the quantizers do not suffer from dynamic range issues.  However, as signs of measurements do not provide amplitude information of the signal, it can be recovered up to a constant scalar factor only. Additionally, more measurements are needed for a successful reconstruction for such systems, exceeding the signal length. In this paper, we  focus on the general $B$-bit quantization of available measurements and its effect to the reconstruction accuracy.

If the reconstruction is performed without measurements quantization, the error will be equal to zero or negligible. The quantization step in measurements inevitably introduces reconstruction errors. The effects of quantization in compressive sensing theory has been recently presented \cite{preprint,Quant1,Quant2,Quant3,Quant4,Bhandari}.  The results mainly include the derivation of quantization error bounds and adaptation of CS algorithms aiming to reduce the distortions related to the quantization \cite{candesQUANT, candesQUANT2}. The upper bound of the reconstruction error, for strictly sparse signals, has been derived in \cite{preprint}. Other reported results are focused on the worst case analysis \cite{Quant1}. Exact asymptotic distortion rate functions have been derived in \cite{Quant1} for scalar quantization, where the reconstruction strategies have been adapted to accommodate quantization errors. An overview of the quantization phenomena in the compressive sensing context is presented in \cite{Quant2}. Therein, the fundamental analysis provides the performance bounds only, with an additional focus on the Sigma-Delta quantization and the related theory. Recently, the effect of quantization on the estimation of sparsity order, support and signals have been studied within a large number of Monte Carlo simulations in \cite{Quant3}. The most frequently used algorithms in compressive sensing are adjusted to the quantization effect in \cite{Quant4}.  For the case of one-bit unlimited sampling, a quantization approach using the one-bit modulo samples,  \cite{Bhandari} shows the bounds of the reconstruction error.

This paper aims to fill the literature gap regarding the exact characterization of the quantization in compressed sensing, by deriving an explicit relation for the mean squared error, instead of the error bounds. The error produced by the quantization of measurements is analyzed from a practical signal processing point of view. Additional to that, the error appearing when approximately sparse signals are reconstructed under the sparsity constraint is also examined. The analysis is expanded to include the effect of the pre-measurements noise in the sparsity domain coefficients, known as the noise folding \cite{Dic5}. The presented theory is unified by exact relations for the expected squared reconstruction errors, derived to take into account all the studied effects. The results are validated using three different reconstruction algorithms. Moreover, we comment on the modifications of the derived relations, required to include the floating-point arithmetics.

The paper is organized as follows. In Section \ref{TradCS}, basic CS concepts and definitions are briefly presented. Section \ref{ProbSol} introduces a common approach to solve the CS reconstruction problem, including also a brief overview of relevant properties which characterize possible solutions.  Section \ref{QuantEffect} puts the quantization within the compressive sensing framework. In Section \ref{NonsparseSec}, the concept of nonsparse (approximately sparse) signals reconstructed under the sparsity constraint is analyzed, leading to the reconstruction error equation which unifies the studied effects. The theory is expanded, to take into account the noise folding effect, in Section \ref{NoiseFoldingSec}, while the Section \ref{FP} discusses the quantization in floating-point arithmetics. Numerical results verify the presented theory in Section \ref{Results}. The paper ends with the concluding remarks. 

\section{Basic Compressive Sensing Definitions} \label{TradCS}

\textit{Definition:} A discrete signal $x(n)$, $n=0,1,\dots,N-1$ is sparse in one of its representation domains $X(k)$ if the number $K$  of nonzero coefficients  is much smaller than the total number of samples $N$, that is,
$$X(k)=0 \textrm{ for } k \notin \mathbb{K}=\{k_1,k_2,\dots,k_K\},$$
where $K \ll N$.

\textit{Definition:} A measurement of a signal is a linear combination of its sparsity domain coefficients  $X(k)$,
\begin{gather}
y(m)=\sum_{k=0}^{N-1} a_m(k)X(k), \,\, m=1,\dots,M,
\end{gather}  
or in matrix form
\begin{gather}
\mathbf{y}=\mathbf{A}\mathbf{X},  
\end{gather}  
where $\mathbf{y}$ is an $M\times 1$ ($M$-dimensional) column vector of the measurements $y(m)$, $\mathbf{A}$ is an $(M\times N)$-dimensional measurement matrix with coefficients $a_m(k)$ as its elements, and $\mathbf{X}$ is an $N\times 1$ ($N$-dimensional) column sparse vector of coefficients $X(k)$. It is common to normalize the measurement matrix such that the energy of its columns is $1$. In that case, the diagonal elements of the matrix $\mathbf{A}^H \mathbf{A}$ are equal to $1$,
 where $\mathbf{A}^H$ denotes a Hermitian transpose of $ \mathbf{A}$. 

By definition, a measurement of a $K$-sparse signal can be written as
$$y(m)=\sum_{i=1}^{K}X(k_i)a_m(k_i).$$


The compressive sensing theory states that, under certain realistic conditions, it is possible to reconstruct a sparse $N$-dimensional vector $\mathbf{X}$ from a reduced $M$-dimensional set of measurements ($M<N$) belonging to vector $\mathbf{y}$,  
\begin{equation}
\mathbf{y}=[y(1),y(2),\dots,y(M)]^T.
\end{equation}


The reconstruction conditions are defined in several forms. The most widely used are the forms based on the restricted isometry property (RIP) and the coherence index \cite{Donoho_CS,Baraniuk_CS,L1_magic,candesQUANT}.  Although providing tight bounds, the RIP based condition is of high calculation complexity. This is the reason why the coherence based relation will be considered in this paper, along with some comments on its probabilistic relaxation. 

\textit{The reconstruction of a $K$-sparse signal, $\mathbf{X}$ is unique if $K < \left(1+1/\mu\right)/2,$ where the coherence index, $\mu$, is equal to the maximum absolute off-diagonal element of $\mathbf{A}^H \mathbf{A}$, assuming its unity diagonal elements.}

A simple proof will be provided later. 

Formally, compressive sensing aims to solve the optimization problem
\begin{align}
\label{L1min}
\text{min  }\left \| \mathbf{X} \right \|_0 \text{ subject to } \mathbf{y=AX},
\end{align}
or its corresponding relaxed convex form. 
Amongst many others, an approach based on matching the components corresponding to the nonzero coefficients, can be used to solve (\ref{L1min}). It is further assumed that the CS reconstruction is based on a such methodology. The solution is discussed in the next section, since it will be used to model the quantization noise and other studied effects.

\section{Problem Solution}
\label{ProbSol}

To perform the reconstruction, we use an iterative version of the orthogonal matching pursuit algorithm from \cite{cosamp}. Assume that $K$ nonzero values $X(k)$ are detected at positions $k \in \mathbb{K}= \{k_1,k_2,\dots,k_K\}$. The system of measurement equations becomes
\begin{equation}
\mathbf{y}=\mathbf{A}_{MK}\mathbf{X}_K.
\end{equation}
The system is solved for the nonzero coefficients $X(k)$, $k \in \mathbb{K}$ written in vector form as $\mathbf{X}_K$. 
The matrix $\mathbf{A}_{MK}$ is an $M \times K$ sub-matrix of the $M \times N$ measurement matrix $\mathbf{A}$, where only the columns corresponding to the positions of nonzero elements in $X(k)$ are kept.  
The solution of the system is 
\begin{equation}
\mathbf{X}_K=(\mathbf{A}^H_{MK}\mathbf{A}_{MK})^{-1}\mathbf{A}_{MK}^H\mathbf{y}=\textrm{pinv}(\mathbf{A}_{MK})\mathbf{y}, \label{pseudo_sol}
\end{equation}
where $\textrm{pinv}(\mathbf{A}_{MK})$ is the pseudo-inverse of the matrix  $\mathbf{A}_{MK}$  and $\mathbf{A}^H_{MK}\mathbf{A}_{MK}$  is known as a $K \times K$ Gram matrix of $\mathbf{A}_{MK}$.  

Therefore, the problem solution can be split into two steps:
\begin{enumerate}
	\item detect the positions of nonzero coefficients, and
	\item apply the reconstruction algorithm at detected positions. 
\end{enumerate} 

\subsection{Initial Estimate}

Detection of the positions of nonzero coefficients $X(k)$ will be based on the initial estimate concept. 
An intuitive idea for the initial estimate comes from the fact that the measurements are obtained as linear combinations of the sparsity domain coefficients, with rows of the measurement matrix $\mathbf{A}$ acting as weights. It means that the back-projection of the measurements $\mathbf{y}$ to the measurement matrix $\mathbf{A}$, defined by
\begin{equation}
\mathbf{X}_0=\mathbf{A}^H \mathbf{y}=\mathbf{A}^H \mathbf{A} \mathbf{X,}\label{Initi}
\end{equation}
can be used to estimate the positions of nonzero coefficients.  
For the coefficient at the $k$th position, its initial estimate $X_0(k)$ takes the following form:
\begin{gather}
X_0(k)=\sum_{m=1}^{M}y(m) a^*_{m}(k), \label{EstLiny}
\end{gather} 
or
\begin{gather}
X_0(k)=\sum_{i=1}^{K}X(k_i) \mu(k_i,k), \label{EstLin}
\end{gather} 
where 
\begin{equation}
\mu(k_i,k)=\sum_{m=1}^{M}a_m(k_i)a^*_{m}(k)
\label{eq:mi}
\end{equation}
are the coefficients of  mutual influence (interference) among elements  $X(k)$. They are equal to the elements of matrix $\mathbf{A}^H \mathbf{A}$, with
\begin{equation}
\mu = \max_{k \ne l} |\mu(l,k)|,
\end{equation}
and $\mu(k,k)=1.$ Note that the $\mu$ represents also the coherence index condition.

For various values of $k_i$, the off-diagonal elements $\mu(k_i,k)$ of matrix $\mathbf{A}^H \mathbf{A}$ act as random variables, with different distribution for different measurement matrices. For the partial discrete Fourier transform (DFT) matrix, distribution of $\mu(k_i,k)$ tends to a Gaussian distribution for $1 \ll M \ll N$, while for an equiangular tight frame (ETF) measurement matrix, $\mu(k_i,k)$ takes only the values such that $|\mu(k_i,k)| = \mu$. Distributions of $\mu(k_i,k)$ for other measurement matrices can be also easily determined. 

The reduced set of measurements (samples) manifests as a noise in the initial estimate, which therefore acts as a random variable, with mean-value and variance given by
\begin{gather}
E\{X_0(k)\}=\sum_{i=1}^{K}X(k_i)\delta(k-k_{i})
\label{mean}
\\
\mathrm{var}\{X_0(k)\}=\sum_{i=1}^{K}|X(k_i)|^{2}\mathrm{var}\{\mu(k_i,k)\}\left(  1-\delta(k-k_{i})\right),
\label{variance}
\end{gather} 
where $\delta(k)=1$ only for $k=0$ and $\delta(k)=0$, elsewhere.

In the analysis of the reconstruction error, we are interested in the variance of random variable $\mu(k_i,k)$, i.e.
$$\mathrm{var}\{\mu(k_i,k)\}=\sigma_{\mu}^2.$$
For the partial DFT matrix, the variance is derived in \cite{Srdjan_CS}. For a real-valued ETF measurement matrix, the values $\pm \mu$ are equally probable, producing the variance $\sigma_{\mu}^2=\mu^2$, where, according to the Welch bound, $\mu^2=(N-M)/(M(N-1))$ holds \cite{Welch,DSPknjiga}.  For the Gaussian measurement matrix, the variance is $\sigma_{\mu}^2=1/M$. The same value is obtained for other considered random matrices. The variance $\sigma_{\mu}^2$  of $\mu(k_i,k)$ is presented in Table \ref{tab:1} for various measurement matrices \cite{Srdjan_CS, DSPknjiga, geosc, tutorial}.

\begin{table}[tb]		
		\centering
		\caption{Variances in the initial estimate for various measurement matrix types for $k \ne k_i$ }
		\label{tab:1}
		\renewcommand{\arraystretch}{1.25}
		\begin{tabular}{lcc}
			\toprule
			\textbf{Measurement matrix} & $a_m(k)$ & $\sigma_{\mu}^2$ \\
			\midrule 
			Partial DFT & $\frac{1}{\sqrt{M}} e^{j2\pi n_m\frac{k}{N}}$ & $\frac{N-M}{M(N-1)}$ \\[3pt]
			Random partial DFT & $\frac{1}{\sqrt{M}} e^{j2\pi t_m\frac{k}{N}}$  & $1/M$ \\[3pt]
			Equiangular tight frame & $\mu=\sqrt{\frac{N-M}{M(N-1)}}$  & $\frac{N-M}{M(N-1)}$ \\[3pt]
			Gaussian random & $\sim \mathcal{N}(0,\frac{1}{M})$ & $1/M$ \\[3pt]
		 	Uniform random & $\sim \mathcal{U}(0,\frac{1}{M})$ & $1/M$ \\[3pt]
			Bernoulli (binary) & $\sim \pm \frac{1}{\sqrt{M}}$ & $1/M$ \\
			\bottomrule 
		\end{tabular}
		\renewcommand{\arraystretch}{1}
\end{table}

\subsection{Detection of Nonzero Element Positions}

The initial estimate can be used as a starting point for an analysis of the reconstruction performance and its outcomes. Potentially, such analysis can lead to the improvements of the reconstruction process. The detection can be done in one step or in an iterative way.

\textbf{One-step detection:} In an ideal case, matrix $\mathbf{A}^H \mathbf{A}$ should be such that the initial estimate $\mathbf{X}_0$ contains $K$ coefficients higher than the other coefficients. Then by taking the positions of the highest coefficients in (\ref{Initi}) as the set $\mathbb{K}$, the signal is simply reconstructed using (\ref{pseudo_sol}).

\textbf{Iterative detection:} The condition that all $K$ nonzero coefficients in  the initial estimate $\mathbf{X}_0$ are larger  than the coefficient values $X_0(k)$ at the original zero-valued positions $k \notin \mathbb{K}$, can be relaxed using an iterative procedure. To find the position of the largest coefficient in $\mathbf{X}$ based on $\mathbf{X}_0$, it is sufficient that the coefficient $X_0(k)$  has a value larger than the values of the coefficients $X_0(k)$ at the original zero-valued coefficient positions  $k \notin \mathbb{K}$.

\begin{Remark}\label{SolUR} 
\textit{Solution uniqueness.} The worst case for the detection of a nonzero coefficient, with a normalized amplitude $1$, occurs when the remaining $K-1$ coefficients are equally strong (i.e. with unity amplitudes). Then, the influence of other nonzero coefficients to the initial estimate of the considered coefficient may assume its highest possible value.   The influence of the $ k$th coefficient on the one at the $i$th position is equal to $\mu(k_i,k)$, given by (\ref{eq:mi}). Its maximum possible absolute value is the coherence index $\mu$. In the worst case, the amplitude of the considered coefficient in the initial estimate is $ 1-(K-1)\mu$.  At the position where there the original coefficient $X(k)$ is zero-valued, 
in the worst case, the maximum possible contributions $\mu$ of all $K$ coefficients sum up \textit{in phase} to produce the maximum possible disturbance $K \mu $. The detection of the strongest coefficient is successful if  
$$ 1-(K-1)\mu > K \mu, $$
producing the well-known coherence condition for the unique reconstruction 
$
K<\left( 1+1/\mu \right)/2. \nonumber 
$ 

After the largest coefficient position is found and its value is estimated, this coefficient can be subtracted and the procedure can be continued with the remaining $(K-1)$-sparse signal. If the reconstruction condition is met for the $K$-sparse signal, then it is met for all lower sparsities as well.  

The procedure is iteratively repeated for each coefficient. The stopping criterion is that $\mathbf{A}_{MK}\mathbf{X}_K=\mathbf{y}$ holds for the estimated positions $\{k_1,k_2,\dots,k_K\}$ and coefficients $X(k)$.

The method is summarized in Algorithm 1.

\end{Remark}


\begin{algorithm}[!h] \label{ReconMet}
	\caption{Reconstruction Algorithm}
	\label{Norm0Alg}
	\begin{algorithmic}[1]
		\Input Vector $\mathbf{y}$, matrix $\mathbf{A}$, assumed sparsity $K$
		\State $\mathbb{K} \gets \emptyset$, $\mathbf{e} \gets \mathbf{y} $
		\For {$i=1$} {$K$}
		\State $k \gets $ position of 
		the highest 
		value in $\mathbf{A}^H\mathbf{e}$
		\smallskip
		\State $\mathbb{K} \gets \mathbb{K} \cup k $
		\State $\mathbf{A}_K \gets $ columns of matrix $\mathbf{A}$ selected by 
		set $\mathbb{K} $
		\State $\mathbf{X}_K \gets \operatorname{pinv}(\mathbf{A}_K)\mathbf{y}$
		\State $\mathbf{y}_K \gets \mathbf{A}_K\mathbf{X}_K$ 
		\State $\mathbf{e} \gets \mathbf{y} -  \mathbf{y}_K$
		\EndFor
		
		\Output
		Reconstructed $\mathbf{X}_R=\mathbf{X}_K$ and positions $\mathbb{K}.$ 
	\end{algorithmic}
\end{algorithm}

\begin{Remark}\label{Soex}
\textit{Solution exactness.} The coherence index condition guarantees that the positions of the nonzero elements in $\mathbf{{X}}$ will be uniquely determined. Next, we will show that the values of nonzero coefficients will be recovered exactly.  

The system of linear equations in (\ref{EstLin}), for $k \in \mathbb{K}$, can be written in a matrix form as
\begin{equation*}
\mathbf{{X}}_{0K}=\mathbf{B} \mathbf{X}_K=\mathbf{X}_K+\mathbf{C}_K,
\end{equation*}
where $\mathbf{B}$ is a $K\times K$ matrix with elements $b_{pi}=\mu(k_p,k_i)$,  $\mathbf{{X}}_{0K}$ is the vector with $K$ elements obtained from the initial estimate as ${X}_{0K}(i)={X}_0(k_i)$, and $\mathbf{X}_K$ is the vector with $K$ corresponding coefficients from the original signal. The influence of the other $K-1$ coefficient to the considered coefficient is denoted by  $\mathbf{C}_K$. 

The reconstructed coefficients $\mathbf{X}_R$, at the nonzero coefficient positions, are obtained by minimizing $\left\| {\mathbf{y} - \mathbf{A}_K\mathbf{X}_R} \right\|_2^2$.  They are
\begin{equation}
\mathbf{X}_R=(\mathbf{A}_K^H \mathbf{A}_K)^{-1} \mathbf{A}_K^H \mathbf{y},
\label{solut}
\end{equation}
where  $\mathbf{A}_K$ is a matrix obtained from measurement matrix $\mathbf{A}$ by keeping the columns for $k \in \mathbb{K}$. Since $\mathbf{A}_K^H \mathbf{y}= \mathbf{{X}}_{0K}$, according to (\ref{Initi}), we can rewrite (\ref{solut}) as
\begin{equation}
\mathbf{X}_R=(\mathbf{A}_K^H \mathbf{A}_K)^{-1}\mathbf{{X}}_{0K}. \end{equation}
Since $\mathbf{{X}}_{0K}=\mathbf{B} \mathbf{X}_K$, the reconstruction is exact if 
$$(\mathbf{A}_K^H \mathbf{A}_K)^{-1}=\mathbf{B}^{-1}$$ 
holds. Indeed, the elements of matrix $\mathbf{A}_K^H
\mathbf{A}_K $ are equal to $\beta_{{i}{j}}=\sum_{n=1}^{M}a^*_n(k_{i}) a_n(k_{j})=\mu(k_{j},k_{i})$ meaning that $\mathbf{A}_K^H
\mathbf{A}_K =\mathbf{B}$.  Therefore,  $\mathbf{X}_R=\mathbf{X}_K$ holds.

The reconstruction algorithm produces correct coefficient values $X(k)$ at the selected positions $k \in \mathbb{K}$. It means that the influence of other $K-1$ coefficients to each coefficient in the initial coefficient estimate ${X}_0(k),$ denoted by $C(k),$ is canceled out. 

\begin{figure}[bt]
	\centering
	\includegraphics[]{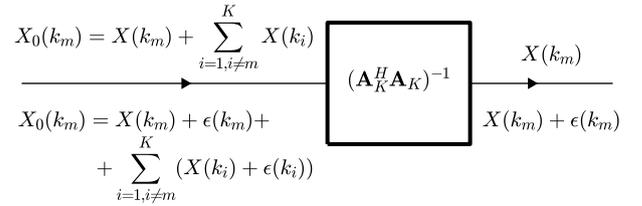}
	\caption{Illustration of a system for the reconstruction of a sparse signal $\mathbf{X}_K=[X(k_1),X(k_2),\dots,X(k_K)]^T$ from the initial estimate $\mathbf{X}_{0K}=[X_0(k_1),X_0(k_2),\dots,X_0(k_K)]^T$.}
	\label{HPassSystem}
\end{figure}

In summary, the reconstruction algorithm for a coefficient at a position $k \in \mathbb{K}$, works as an identity system to the original signal coefficient in $ X_0(k),$ eliminating the influence of other coefficient at the same time, Fig. \ref{HPassSystem}. 
\end{Remark}

\subsection{Noisy Measurements}

Assume next that the observations are noisy
\begin{equation}
	\mathbf{y}+\boldsymbol{\varepsilon}=\mathbf{AX},
\end{equation}
with a zero-mean signal independent noise $\boldsymbol{\varepsilon}$. The noise variance of the assumed additive input noise $\boldsymbol{\varepsilon}$ is $\sigma^2_{\varepsilon}$ and the covariance is given by $$\mathrm{E}\{\boldsymbol{\varepsilon}\boldsymbol{\varepsilon}^H\}=\sigma_{\varepsilon}^2\mathbf{I}.$$
 Noisy measurements will result in a noisy estimate  $\mathbf{X}_0=\mathbf{A}^H (\mathbf{y}+\boldsymbol{\varepsilon})$. Variance of $X_0(k)$ due to the input noise in measurements, is  $\sigma^2_{{X}_0(k)}=\sigma^2_{\varepsilon}$ since it has been assumed that the columns of $\mathbf{A}$ have unite energy,
 $$\mathrm{E}\{\mathbf{X}_0\mathbf{X}^H_0\}-|\mathrm{E}\{\mathbf{X}_0\}|^2=\mathrm{E}\{\mathbf{A}^H\boldsymbol{\varepsilon}\boldsymbol{\varepsilon}^H\mathbf{A}\}=\sigma_{\varepsilon}^2\mathbf{I}.$$
The noise variance in the reconstructed coefficient is (Remark \ref{Soex} and Fig. \ref{HPassSystem})
$$\operatorname{var} \{X_R(k)\}=\sigma^2_{\varepsilon}.$$

Since the noise is independent in each reconstructed coefficient, the total mean squared error (MSE) in $K$ reconstructed coefficients is 
\begin{align}
\left\| {{\mathbf{X}_R} - {\mathbf{X}_K}} \right\|_2^2 = K{\sigma^2_\varepsilon }.
\label{sumSam}
\end{align}

If the partial DFT matrix is formed as a submatrix of the standard inverse DFT matrix (with normalization $1/N$), then we would get $\left\| {{\mathbf{X}_R} - {\mathbf{X}_K}} \right\|_2^2 = KN^2{\sigma _\varepsilon }^2/M$, as shown, for example, in \cite{geosc}.

\section{Quantization Effects}
\label{QuantEffect}

Traditional CS theory does not consider the limitations in the number of bits used for the measurements representation. This can affect the reconstruction performance of the standard CS approaches. 

The measurement quantization is particularly important in the hardware implementation context. One-bit measurements are the most extreme case, promising simple, comparator-based hardware devices \cite{1bitCS}. The one bit used represents the sign of the sample, i.e.  $\mathbf{y}=\text{sign}\{\mathbf{AX}\}$. However, a larger number of samples is required for an accurate reconstruction, which is difficult to achieve using only the sign of a measurement \cite{1bitCS,BIHTpaper}.

 A more general form of the hardware implementation uses a $B$-bit digital sample of a measurement. We will assume that the measurements are stored into $(B+1)$-bit registers (one sign bit and $B$ bits for the signal absolute value),
\begin{equation}
\mathbf{y}_B=\text{digital}_B\{\mathbf{AX}\},
\end{equation}
whereas the reconstruction of the coefficients $X(k)$ is done in a more realistic sense for hardware purposes. The requirement for storage is also significantly reduced for such measurements, since the total number of bits is reduced. Note that, for a complex-valued signal $x(n)$, the measurements $\mathbf{y}_B$ are also complex, formed as
\begin{equation}
\mathbf{y}_B=\text{digital}_B\{\Re\{\mathbf{AX}\}\}+j\text{digital}_B\{\Im\{\mathbf{AX}\}\}
\end{equation}
where both real and imaginary part of measurements are quantized to $B$-bits.

\subsection{Quantization errors}
Quantization influences the results of the compressive sensing reconstruction in several ways:

\begin{itemize}
\item Input signal quantization error, described by an additive quantization noise.
This  influence can be modeled as a uniform noise with values between the quantization level bounds.

\item Quantization of the results of arithmetic operations. It depends on the way how the calculations are performed.

\item Quantization of the coefficients in the algorithm. However, being deterministic for a given measurement matrix, this type of error is commonly neglected from the analysis. 
\end{itemize}

In order to perform an appropriate and exact analysis, some standard assumptions are further made:

\begin{itemize} 
	
	\item The quantization error is a white noise process with a uniform distribution. 
	\item The quantization errors are uncorrelated.
	
\item The quantization errors are not correlated with the input signal.

\end{itemize}

The most important source of an error is the quantization of the measurements $y(m)$ and the quantization of the measured sparse signals $X(k)$, referred to as the quantization noise folding. They will be analyzed next.


\subsection{Input signal ranges}

Assume that registers with $B$ bits, with an additional sign bit, are used and that all measurements are
normalized to the range 
$$ -1 \le y(m)< 1.$$ 
The total number of bits in a register is $b=B+1$.

In that case, it is important to notice that the sparse signal coefficients $X(k)$ must be within the range $ -\min\{\sqrt{M}/K,1\} < X(k)< \min\{\sqrt{M}/K,1\}$ so that $\mathbf{y}=\mathbf{A}_{MK}\mathbf{X}_K$ does not produce a value with amplitude greater or equal to $1$. For the partial DFT matrices, this condition holds in a strict sense, while for the Gaussian matrices it holds in a mean sense (all values whose amplitudes are greater than $1$ are quantized to the closest level with amplitude below $1$). Note that the butterfly schemes for the measurements calculation (as in the quantized FFT algorithms) could extend this bounds for $X(k)$ so that the maximum range $ -1 < X(k)< 1$ can be used.   

\subsection{Measurements quantization}

For the $B$-bit registers, the digital signal values $\mathbf{y}_B$ are coded
into a binary format. When the signal amplitude is quantized to $B$ bits, the difference in amplitude which produces the quantization is called the quantization error. The quantization error is bounded by
\begin{equation}
|e(m)| < \Delta/2,
\end{equation}
where $\Delta$ is related to $B$ through
\begin{equation}
\Delta = 2^{-B}.
\end{equation}

The quantization error of a signal can be defined as an additive uniform white noise affecting the measurements
\begin{equation}
\mathbf{y} = \mathbf{y}_B + \mathbf{e},
\end{equation}
where $\mathbf{e}$ is the quantization error vector with elements $e(m)$.
The mean and variance of the quantization noise are calculated as \cite{DSPknjiga}
\begin{gather}
\mu_\mathbf{e} = \mathrm{E}\{\mathbf{e}\} = 0,\\
\sigma^2_\mathbf{e} = \Delta^2/12.
\label{varReal}
\end{gather} 
Note that, for a complex-valued signal, both real and imaginary parts of samples contribute to the noise. Therefore, in this case, the variance of the quantization noise can be written as
\begin{equation}
\sigma^2_\mathbf{e} = 2\Delta^2/12 = \Delta^2/6.
\end{equation}

Considering $\mathbf{y}$ as noisy measurements, the initial estimate will result in a noisy  $X_0(k)$. Since $X_0(k)$ is calculated from (\ref{EstLiny}), with the quantization noise in measurements, its variance will be 
\begin{equation}
\sigma^2_{X_0(k)}=\sigma^2_{\mathbf{e}}.
\end{equation}


 Therefore, the noise variance in the output (reconstructed) coefficients, for the system shown in Fig. \ref{HPassSystem}, is equal to the input noise variance
\begin{equation}
\operatorname{var} \{X_R(k)\}=\sigma^2_{\mathbf{e}}.
\end{equation}

Since only $K$ out of $N$ coefficients are used in the reconstruction, the energy of the reconstruction error is
\begin{align}
\left\| {{\mathbf{X}_R} - {\mathbf{X}_K}} \right\|_2^2 = K{\sigma^2 _\mathbf{e} },
\label{sumSam}
\end{align}
where for notation simplicity we have used $\left\| {{\mathbf{X}_R} - {\mathbf{X}_K}} \right\|_2^2$ to denote the expected value of the squared norm-two of the vector ${\mathbf{X}_R} - {\mathbf{X}_K}$. The full and complete notation of the left side of (\ref{sumSam}) would be $\mathrm{E}\{\left\| {{\mathbf{X}_R} - {\mathbf{X}_K}} \right\|_2^2\}$.

\subsection{Sparsity to Number of Bits Relation}

Based on the previous relations, influence of the quantization with $B$ bits can be related to the sparsity $K$. The error energy in the reconstructed coefficients will remain the same if
\begin{equation}
K\sigma^2 _\mathbf{e}=K\frac{2^{-2B}}{6} =\text{const.}
\end{equation}
It means that reduction of $B$ to $B-1$ will require sparsity reduction from $K$ to $K/4$.  The logarithmic form of the reconstruction error is
\begin{gather*}
e^2=10\log \big( \left\| {{\mathbf{X}_R} - {\mathbf{X}_K}} \right\|_2^2 \big) 
=  3.01 \log_2 K - 6.02 B-7.78.
\label{sumSamlog}
\end{gather*}


\section{Nonsparsity Influence}
\label{NonsparseSec}

Due to many circumstances, majority of signals in real-world scenarios are only approximately sparse or nonsparse. This means that a signal, in addition to the $K$-sparse large coefficients,  has $N-K$ coefficients in the sparsity domain which are small but nonzero. Assume such an approximately sparse (or nonsparse) signal $\mathbf{X}$. The signal is reconstructed under the $K$-sparsity constraint using Algorithm 1, with the reconstruction conditions being satisfied in the compressive sensing sense, thus allowing that the algorithm can detect the $K$ largest coefficients.

The reconstructed signal $\mathbf{X}_{R}$ then has $K$ reconstructed coefficients with amplitudes $X_R({k_1}), \  X_R({k_2}), \dots, X_R({k_K})$. The remaining $N-K$ coefficients which are not reconstructed are treated as a noise in these $K$ largest coefficients. Variance from a nonzero coefficient, according to (\ref{variance}), is $\left\vert X(k_l)\right\vert ^{2}(N-M)/(M(N-1)).$
The total energy of noise in the $K$ reconstructed coefficients
$\mathbf{X}_{R}$ will be
\begin{equation}
\left\Vert \mathbf{X}_{R}\mathbf{-X}_{K}\right\Vert _{2}^{2}=K \sigma_{\mu}^2\sum_{l=K+1}^{N}\left\vert X(k_l)\right\vert ^{2},
\end{equation}
where $\mathbf{X}_K$ is the sparse version of the original (nonsparse) signal, i.e. a signal with $K$ largest coefficients from  $\mathbf{X}$, and others set to zero. Denoting the energy of remaining signal, when the $K$ largest coefficients are removed from the original signal, by
\begin{equation}
\left\Vert \mathbf{X-X}_{K}\right\Vert _{2}^{2}=\sum_{l=K+1}^{N}\left\vert X(k_l)\right\vert ^{2}
\end{equation}
we get
\begin{equation}
\left\Vert \mathbf{X}_{R}\mathbf{-X}_{K}\right\Vert _{2}^{2}=K\sigma_{\mu}^2 \left\Vert \mathbf{X-X}_{K}\right\Vert _{2}^{2}.
\label{Nonsparse}
\end{equation}

For the partial DFT measurement matrix, the result will be 
\begin{equation}
\left\Vert \mathbf{X}_{R}\mathbf{-X}_{K}\right\Vert _{2}^{2}= K\frac{N-M}
{M(N-1)}\left\Vert \mathbf{X-X}_{K}\right\Vert _{2}^{2}.
\label{NonsparseDFT}
\end{equation}

In the case when the signal is strictly $K$-sparse, i.e. $\mathbf{X=X}_{K}$,  and if the reconstruction is performed with non-quantized measurements, the reconstruction would be ideal and the error would be  $\left\Vert \mathbf{X}_{R}\mathbf{-X}_{K}\right\Vert _{2}^{2}=0$ (or negligible). Since the measurements are quantized by $B$-bits, error of the form (\ref{sumSam}) will be introduced.  

In the case of a nonsparse signal, a general expression is obtained by combining (\ref{sumSam}) and (\ref{Nonsparse}) to get
\begin{equation}
\left\Vert \mathbf{X}_{R}\mathbf{-X}_{K}\right\Vert _{2}^{2}=K\sigma_{\mu}^2 \left\Vert \mathbf{X-X}_{K}\right\Vert _{2}^{2}+K\sigma_{\mathbf{e}}^{2}.
\end{equation}

This result will be validated by examples in the next section, by calculating the signal-to-noise ratio (SNR) of each result 
\begin{equation}
SNR_{th} = 10\log \Bigg( \frac{\left\Vert \mathbf{X}_{K}\right\Vert _{2}^{2}}{K\sigma_{\mu}^2\left\Vert \mathbf{X-X}_{K}\right\Vert _{2}^{2}+K\sigma
_{\mathbf{e}}^{2}.} \Bigg) \label{SNR2}
\end{equation}
and comparing it with the statistical SNR given by
\begin{equation}
SNR_{st} = 10\log \Bigg( \frac{\left\Vert \mathbf{X}_{K}\right\Vert _{2}^{2}}{\left\Vert \mathbf{X}_{R}\mathbf{-X}_{K}\right\Vert _{2}^{2}} \Bigg).  \label{SNR1}
\end{equation}

\section{Noise Folding Quantization}
\label{NoiseFoldingSec}

Additionally to the nonsparse case, we extend our analysis to include the case when a quantization noise $\mathbf{z}$ exists in the signal coefficients $\mathbf{X}$, prior to taking the measurements \cite{Dic5}. In this case, the measurements are of the form 
\begin{equation}
\mathbf{y}_B+\mathbf{e}=\mathbf{A(X+\mathbf{z})},
\label{fold_mod}
\end{equation}
which can be rewritten as
\begin{equation}
\mathbf{y}_B+\mathbf{v=AX}\label{fold_mod2}
\end{equation}
where $\mathbf{v}=\mathbf{e}-\mathbf{Az}$, and the total quantization noise affecting the signal measurements is denoted by $\mathbf{e}$ with covariance $\sigma_{\mathbf{e}}^2\mathbf{I}.$ The quantization noise vector $\mathbf{z}$ is random with covariance $\sigma_{\mathbf{z}}^2\mathbf{I}$, being independent of $\mathbf{e}$. Therefore, the resulting noise $\mathbf{v}$ is characterized by a covariance matrix 
\begin{equation}
\mathbf{C}=\sigma_{\mathbf{e}}^2\mathbf{I}+\sigma_{\mathbf{z}}^2\mathbf{AA}^H.
\end{equation}
If the considered  measurement matrix $\mathbf{A}$ is formed as a partial Fourier matrix,  the relation  $\mathbf{AA}^H=\frac{N}{M}\mathbf{I}$ holds.  The variance of $\mathbf{v}$ is then  
\begin{equation}
\sigma_{\mathbf{v}}^2=\sigma_{\mathbf{e}}^2+\frac{N}{M}\sigma_{\mathbf{z}}^2, 
\end{equation}
with the covariance matrix $\mathbf{C}=\sigma_{\mathbf{v}}^2\mathbf{I}$. 

However, for the sparse case, the quantization error is present in only $K$ nonzero elements of $\mathbf{X}$. It means that the noise $\mathbf{A}\mathbf{z}$ variance is $\frac{K}{M}\sigma_{\mathbf{e}}^2$ or 
\begin{align}
\left\Vert \mathbf{X}_{R}-\mathbf{X}_{K}\right\Vert _{2}^{2}=K\sigma_{\mathbf{e}}^{2}+\frac{K}{M}\sigma_{\mathbf{z}}^{2}.
\label{noiseFoldSparse}
\end{align}

For the nonsparse partial DFT matrix case, $K(N-M)/(M(N-1))\left\Vert \mathbf{X-X}_{K}\right\Vert _{2}^{2}$ is added to the right part of the Eq. (\ref{noiseFoldSparse})
\begin{align}
\left\Vert \mathbf{X}_{R}-\mathbf{X}_{K}\right\Vert _{2}^{2}=K\sigma_{\mathbf{e}}^{2}+\frac{K}{M}\sigma_{\mathbf{z}}^{2}+K\frac{N-M}{M(N-1)}\left\Vert \mathbf{X-X}_{K}\right\Vert _{2}^{2},
\label{noiseFoldNonSparse}
\end{align}
 where it is assumed that the quantization of $K$ largest elements in $\mathbf{X}$ is dominant in that part of error.
  It is shown that all previous relations, for various measurement matrices, can be applied to this case.  

\section{Floating Point Registers}
\label{FP}
\smallskip
In floating point registers, the quantization error is modeled as a multiplicative error
\begin{equation}
y_B(m) = y(m)+ y(m)e(m),
\end{equation}
where $\mathbf{e}$ is the quantization error vector with elements $e(n)$.
As in the classical digital signal processing, for the analysis of floating point arithmetics, it will be assumed that the sparse coefficients $X(k_i)$, $i=1,2,\dots,K$, are independent, equally distributed zero-mean random variables, with variance $\sigma^2_X$.   The coefficients $X(k_i)$ are statistically independent from the measurement matrix $\mathbf{A}$ elements $a_m(k)$.  The mean value of the quantization error is    
$$\mathrm{E}\{y(m)e(m)\}=0.$$
The variance is
$$\mathrm{E}\{y^2(m)e^2(m)\}=\mathrm{E}\{y^2(m)\}\mathrm{E}\{e^2(m)\}.$$
For $y(m)=\sum_{i=1}^{K}X(k_i)a_m(k_i)$ we get
$$\mathrm{E}\{y^2(m)\}=\sigma^2_X\sum_{i=1}^{K}\mathrm{E}\{|a_m(k_i)|^2\}=\sigma^2_X \frac{K}{M},$$
for all measurement matrices with normalized energy columns, when their elements $a_m(k)$ are equally distributed.  
This means that the quantization noise $y(m)e(m)$ has the variance $\sigma^2_X\sigma^2_{\mathbf{e}}K/M$ and we can write (Remark \ref{Soex} and Fig. \ref{HPassSystem})
\begin{equation}
\sigma^2_{X_0(k)}=\frac{K}{M} \sigma^2_X \sigma^2_{\mathbf{e}}
\end{equation}
with 
\begin{align}
\left\| {{\mathbf{X}_R} - {\mathbf{X}_K}} \right\|_2^2 =  \frac{K^2}{M}\sigma^2_X\sigma^2_{\mathbf{e}}.
\label{sumSamFLP}
\end{align}
All formulas, in various considered scenarios, can now be rewritten, including the cases of nonsparse signals and noise folding.  

For example, if the measurements are normalized such  that $\mathrm{E}\{y^2(m)\}=\sigma^2_X K/M=1$ then $\left\| {{\mathbf{X}_R} - {\mathbf{X}_K}} \right\|_2^2 =  K\sigma^2_{\mathbf{e}},$ that is the floating-point arithmetics produces the same results as the fixed-point arithmetics. However, if the range of measurement values is lower, for example, $\mathrm{E}\{y^2(m)\}=\sigma^2_X K/M=1/10$, then the floating-point arithmetics will produce ten times lower error, $\left\| \mathbf{X}_R - \mathbf{X}_K \right\| _2^2 =  K\sigma^2_{\mathbf{e}}/10$.

\section{Numerical Results}
\label{Results}
\smallskip

\textbf{Example 1}: One realization of a sparse and nonsparse signal will be considered as an illustration of the reconstruction. 

\noindent a) Consider an $N=256$-dimensional signal of sparsity $K=10$, whose $M=N/2$ available measurements are stored into registers with $B=6$ bits. The measurements matrix is a partial DFT matrix with randomly selected $M$ out of $N$ rows from the full DFT matrix, with columns being energy normalized. The the sparsity domain coefficients are assumed in the form
\begin{equation}
X(k_p)=\left\{ 
\begin{array}{ll}
\frac{\sqrt{M}}{K}(1- \nu(p)), & \text{for~} p=1,\dots,K\\
0, & \text{for~}  p=K+1,\dots,N, \\
\end{array} 
\right.
\label{sparseSig}
\end{equation}
where $\nu(p)$ is a random variable with uniform distribution from $0$ to $0.4$. 

\begin{figure}
\centering
\includegraphics[]{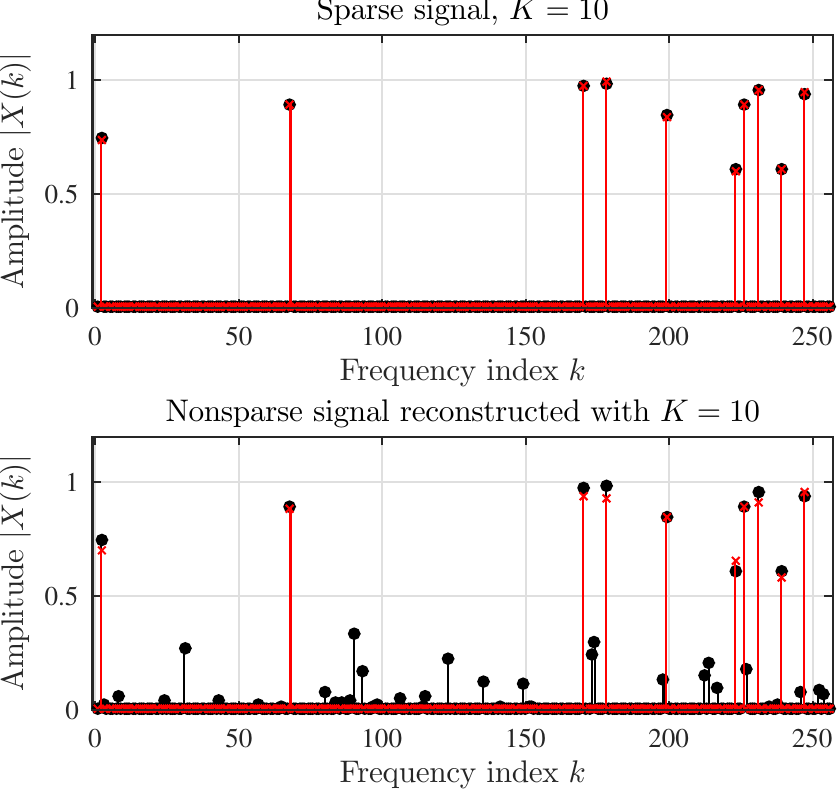}
\caption{Reconstruction results for $N=256$-dimensional signal whose $M=128$ measurements are stored into $B=6$ bit registers. Reconstruction of a sparse signal with $K=10$ nonzero coefficients (top). Reconstruction of a nonsparse signal assuming its sparsity $K=10$ (bottom). The original signal is colored in black, while the reconstructed signal is denoted by red crosses.}
\label{Examp1}
\end{figure}

Since this signal is sparse, the reconstruction error is defined by (\ref{sumSam}). The SNR is defined by (\ref{SNR2}) with $\left\Vert \mathbf{X-X}_{K}\right\Vert _{2}^{2}=0$. The original and the reconstructed signals are shown in Fig. \ref{Examp1}(top). The statistical SNR is $SNR_{st} = 42.35$ dB and $SNR_{th} = 42.56$ dB.

\bigskip
\noindent b)  The signal from a), with $K=10$ significant coefficients, is considered here. However, we will also assume that the remaining $N-K$ coefficients are small but not zero-valued,
\begin{equation}
X(k_p)=\left\{ 
\begin{array}{ll}
\frac{\sqrt{M}}{K}(1- \nu(p)), & \text{for~} p=1,\dots,K \\[5pt]
\frac{\sqrt{M}}{K}\exp(-p/K), & \text{for~} p=K+1,\dots,N, \\
\end{array} 
\right.
\label{nonsparseSig}
\end{equation}
with $\nu(p)$ being a random variable with uniform distribution from $0$ to $0.4$ as in a)  and $N=256$. The number of bits in the registers where the measurements are stored is $B=6$. The original signal and the signal reconstructed under sparsity constraint, using  $M=N/2$ measurements,  are shown in Fig. \ref{Examp1}(bottom). The statistical SNR is $SNR_{st} = 33.33$ dB and the theoretical is $SNR_{th} = 33.68$ dB.

\begin{figure*}
	\centering
	\includegraphics{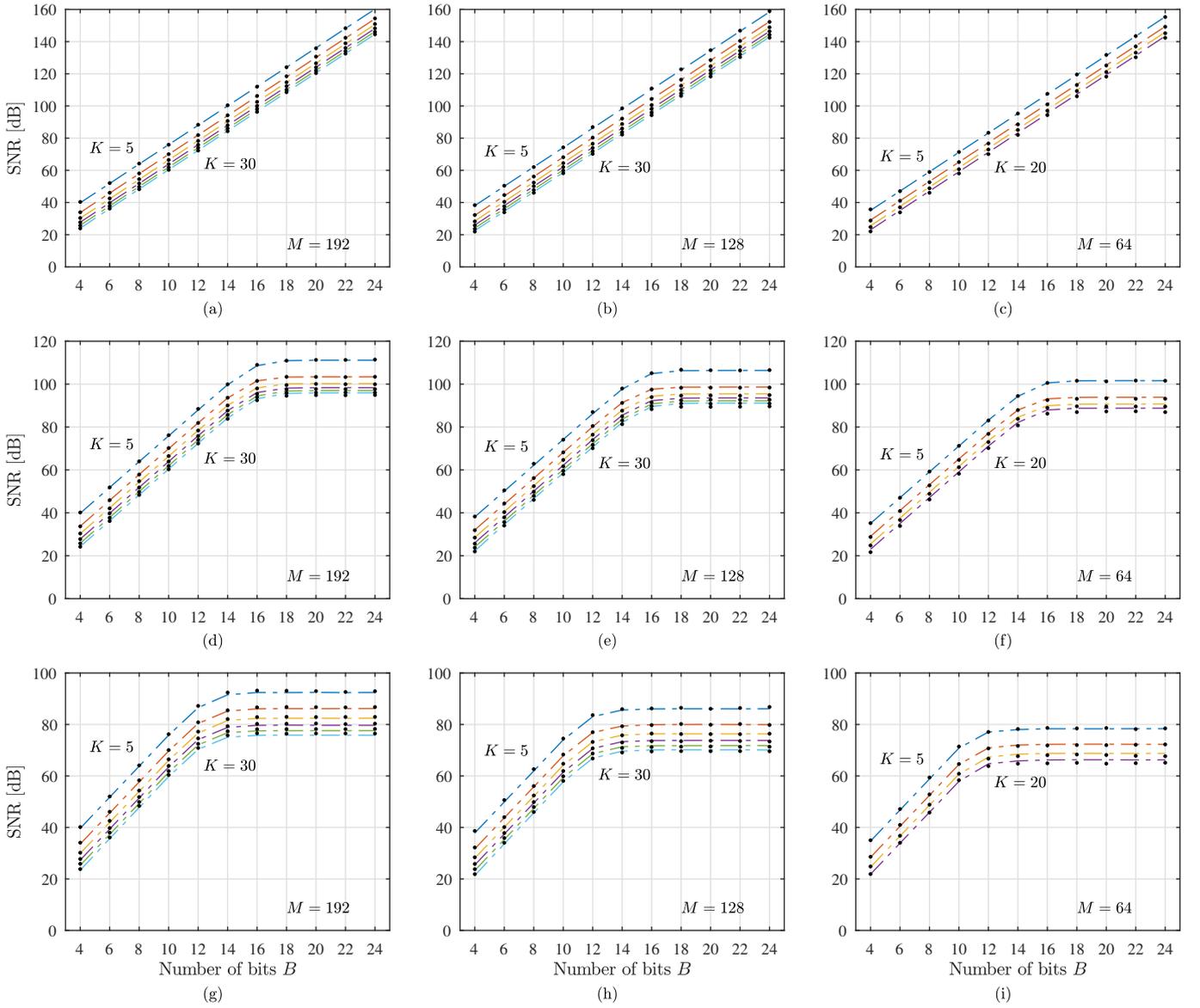}
	\caption{Reconstruction error with measurements quantized to fit the registers with $B$ bits for various sparsities and the numbers of measurements. Partial DFT measurement matrix is used. Statistical results are marked by black dots, while the theoretical results are shown by dot-dashed lines. Sparsities are varied from $K=5$ to maximum $K$ indicated in panels, with a step of $5$. (a)-(c) Reconstruction error (theory and statistics) for sparse signals when only the measurements $\mathbf{y}$ are quantized to $B$ bit fixed-point registers. (d)-(f) Reconstruction error (theory and statistics) for nonsparse signals when only the measurements $\mathbf{y}$ are quantized to $B$ bit fixed-point registers. (g)-(i) Reconstruction error (theory and statistics) for nonsparse signals when both the measurements $\mathbf{y}$ and noisy input coefficients $\mathbf{X}$ are quantized to $B$ bit fixed point registers (quantization noise folding with additive input noise). }
	
	\label{DigOMP_DFT}
\end{figure*}

\textbf{Example 2}: Statistical analysis of the sparse signal reconstruction  whose form is given in (\ref{sparseSig}) is performed in this example. Random uniform changes of coefficient amplitudes $\nu(p)$ are assumed from $0$ to $0.2$.  The numbers of quantized measurements $M=192$, $M=170$, and $M=128$ are considered. Typical cases for the measurements quantization to $B\in \{4,6,8,10,12,14,16,18,20,24\}$ bits are analyzed. 

Signals with sparsity levels $K \in \{5,10,15,20,25,30\}$ are considered. The average statistical signal-to-nose ration $SNR_{st}$ and theoretical signal-to-noise ration $SNR_{th}$ values in 100 realizations are presented in Fig. \ref{DigOMP_DFT}(a)-(c). Black dots represent the statistical results, $SNR_{st}$, and the dash-dot lines show the theoretical results, $SNR_{th}$. The agreement is high. 

For nonsparse signals we used the model in (\ref{nonsparseSig}). Random changes of coefficient amplitudes $\nu(p)$ are assumed from $0$ to $0.2$, while the amplitudes of the coefficients $X(k)$ for $k_p \notin \mathbb{K}$ are of the form $X(k_p)=\exp(-p/(8K))$ in order to reduce its influence to the quantization level. With such amplitudes of the nonsparse coefficients, the quantization error dominates in the reconstruction up to $B=14$, while the nonsparse energy is dominant for $B \ge 16$, as can be seen in Fig. \ref{DigOMP_DFT}(d)-(f). Statistics is again in full agreement with the theoretical results.  

Finally, the noise folding effect is included, taking into account that the input coefficients $X(k)$ are quantized, in addition to the quantization of measurements $y(m)$. Since the folding part of quantization error is multiplied by $K/M \ll 1$ in (\ref{noiseFoldSparse}), the results do not differ from those presented in Fig. \ref{DigOMP_DFT}(a)-(c). In order to test the influence of noise folding we assumed that the quantized input coefficients $X(k)$ contain an additional  noisy. An input additive complex-valued i.i.d. Gaussian noise  with variance $\sigma_z=0.0001$ is added to these coefficients. This noise is of such a level that it does not influence quantization error for $B<14$. However, for $B\geq 14$, it becomes larger than the quantization error and its influence becomes dominant. The results with the quantization and noise folding, with additional noise, are shown in Fig. \ref{DigOMP_DFT}(g)-(i).

\begin{figure*}
	\centering
		\includegraphics{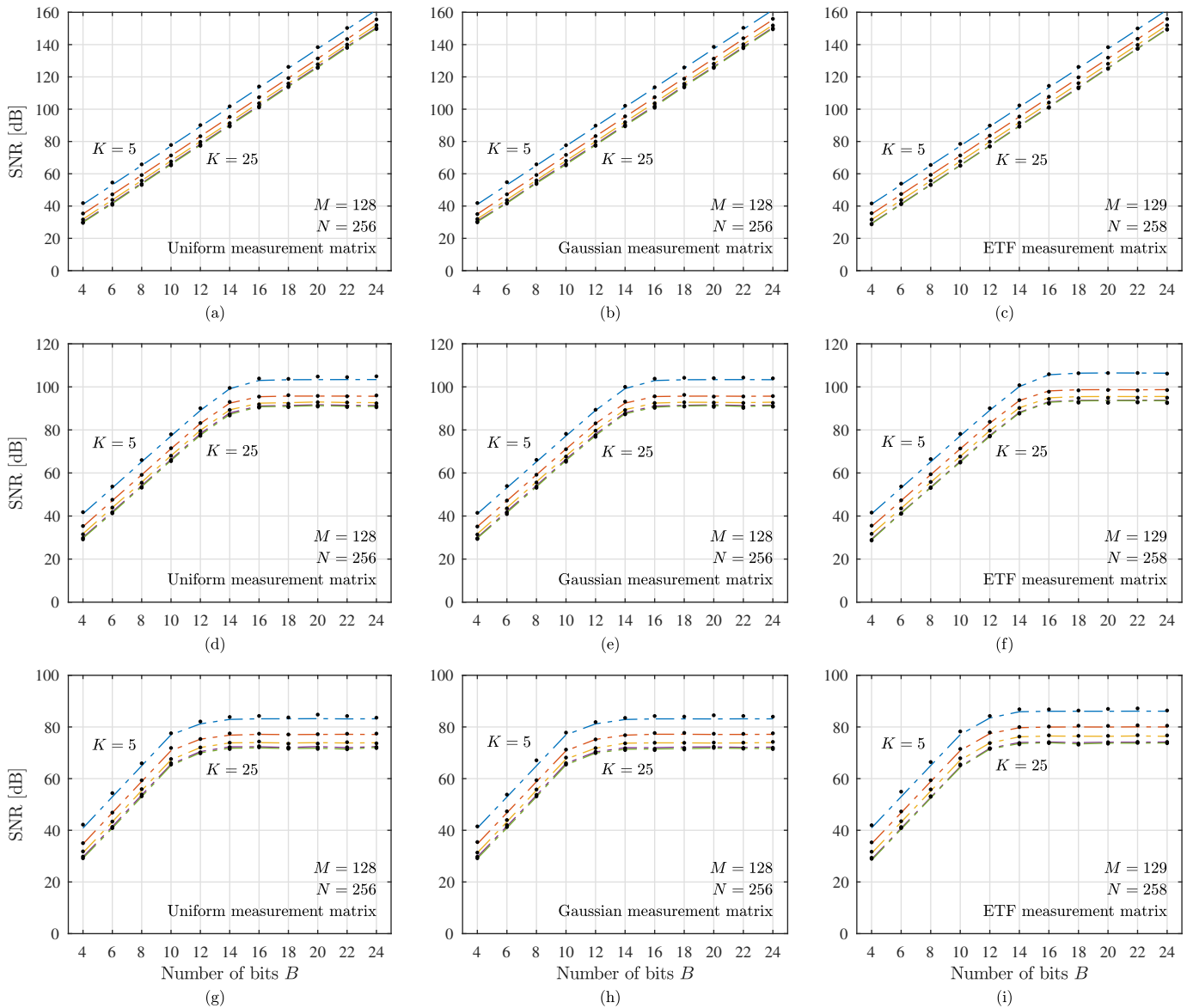}
	\caption{Reconstruction error for various measurement matrices. (a)-(c) Sparse signal with measurements quantized to fit the registers with $B$ bits for various sparsities using a	uniform, Gaussian, and ETF measurement matrices, respectively. (d)-(f) Nonsparse signal with measurements quantized to fit the registers with $B$ bits for various sparsities using a	uniform, Gaussian, and ETF measurement matrices, respectively. (g)-(i)  Nonsparse signals when both the measurements $\mathbf{y}$ and input coefficients $\mathbf{X}$ are quantized to $B$ bit fixed point registers (quantization noise folding)using a	uniform, Gaussian, and ETF measurement matrices, respectively.}
	\label{DigOMP_ostale}
\end{figure*}

\bigskip 
\textbf{Example 3:} The statistical analysis is extended to other forms of the measurement matrices, namely the ETF, the Gaussian, and the uniform random matrix. All three forms of the signal and quatization error are considered here with $M=128$ measurements. Sparse and nonsparse signals described in Example 2 are used in the analysis. The reconstruction error with various number of bits $B=\{4,6,8,10,12,14,16,18,20,24\}$ used in the quantization, and various assumed sparsity levels $K$ are shown in Fig. \ref{DigOMP_ostale}(a)-(c). The results for nonsparse signals, reconstructed with assumed sparsity, are presented in Fig. \ref{DigOMP_ostale}(d)-(f). The noise folding is analyzed for a reduced number of bits in the quantization of $X(k)$ and  presented in Fig. \ref{DigOMP_ostale}(g)-(i).

\medskip 
\begin{figure}[tb]
	\centering
		\includegraphics{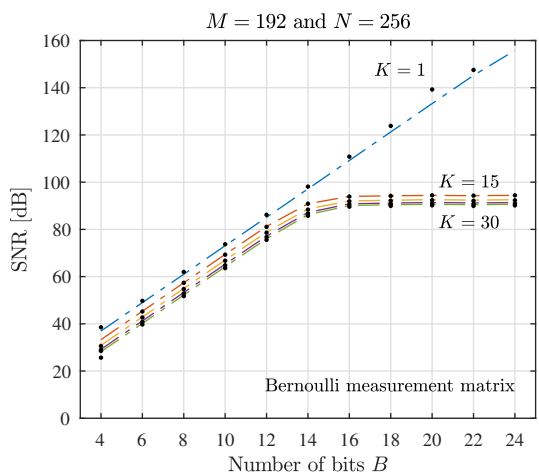}
	\caption{Reconstruction error for the Bernoulli matrix for nonsparse signals with measurements quantized to fit the registers with $B$ bits for various sparsities $K \in \{1,15,20,25,30\}.$ }
	\label{DigOMP_bern_N256_M192_S1_F0}
\end{figure}

\begin{figure*}[ptb]
	\centering
		\includegraphics{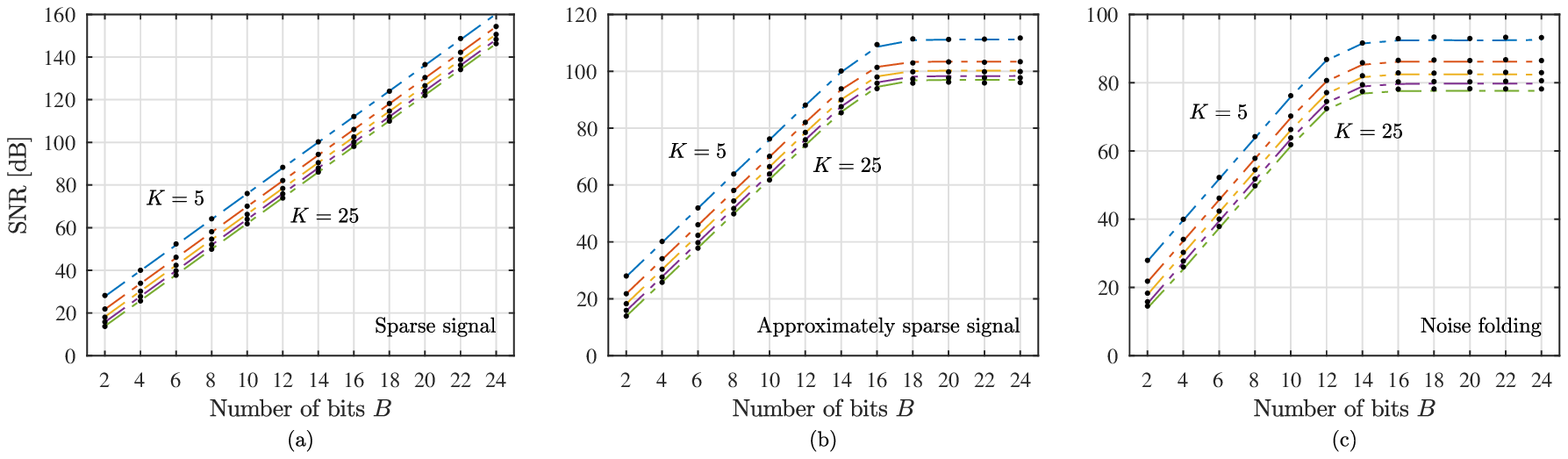}
	\caption{Reconstruction error for the partial DFT measurement matrix when the Iterative Hard Thresholding (IHT) reconstruction algorithm is used. (a) Sparse signal with measurements quantized to fit the registers with $B$ bits for various sparsities. (b) Nonsparse signal with measurements quantized to fit the registers with $B$ bits for various sparsities. (c)  Nonsparse signals when both the measurements $\mathbf{y}$ and input coefficients $\mathbf{X}$ are quantized to $B$ bit fixed point registers (quantization noise folding).}	
	\label{DigHT_OneBit}
\end{figure*}

\textbf{Example 4:} The analysis of quantization effects is done with the assumption that the quantization errors are uncorrelated. This condition is met for all previously considered matrices. However, in the case of Bernoulli measurement matrix and a small signal sparsity this condition does not hold, meaning that we can not expect quite accurate estimation of the statistical error using the previous formulas. To explain this effect, we will start with the simplest case of signal whose sparsity is $K=1$. The measurements are $y(m)=a_m(k_1)X(k_1)$. For all previously considered matrices $y(m)$ and $y(n)$ are different for $m\ne n$ and the quantization errors are independent. However, for the Bernoulli measurement matrix we have $y(m)=\pm X(k_1)/\sqrt{M}$. These measurements will produce only two possible quantization errors for all $m=1,2,\dots,M$. It means that $M/2$ errors in the initial estimate will sum up in phase, producing the mean square error with variance $\operatorname{var} \{X_R(k)\}=\frac{M}{2}\sigma^2_{\mathbf{e}}$. This is significantly higher than $\operatorname{var} \{X_R(k)\}=\sigma^2_{\mathbf{e}}$ in other cases. For $K=2$, we get the measurements $y(m)=(\pm X(k_1) \pm X(k_2))/\sqrt{M}$, producing four possible values for $y(m)$ and only four possible values of the  quantization error. For large $K$, the number of possible levels increases and the result for the variance converges to the one for uncorrelated quantization errors $\operatorname{var} \{X_R(k)\}=\sigma^2_{\mathbf{e}}$, obtained under the assumption that all $M$ measurements $y(m)$ are different.    
The results for the Bernoulli measurement matrix, with the described correstion for small $K$, are shown in Fig. \ref{DigOMP_bern_N256_M192_S1_F0}.

\medskip 
\textbf{Example 5:} The result in previous four examples are obtained based on the OMP reconstruction method (Algorithm \ref{Norm0Alg}), which is also used for the derivation of the theoretical results. Here we will show that we may expect similar results for other reconstruction methods as far as the reconstruction conditions are met. The simulation for the reconstruction with the partial DFT measurement matrix with sparse and nonsparse signals, including noise folding, are repeated with the iterative hard thresholding (IHT) reconstruction method, given in Algorithm \ref{Norm0AlgIHT} \cite{IHTpaper,Quant4}. The theoretical and statistical errors are shown in Fig. \ref{DigHT_OneBit}(a)-(c), showing high agreement between the statistical and theoretical results.

\begin{algorithm}[tbh] \label{ReconMet}
	\caption{Iterative Hard Thresholding (IHT) Reconstruction Algorithm}
	\label{Norm0AlgIHT}
	\begin{algorithmic}[1]
		\Input Vector $\mathbf{y}$,  Matrix $\mathbf{A}$, Assumed sparsity $K$, 
		
		Number of iterations $I_t$, and parameter $\tau$.
		\State $\mathbf{X}_0\gets \mathbf{0}$
		\For {$i=1$} {$I_t$}
		\State $\mathbf{Y} \gets \mathbf{X}_0+ \tau \mathbf{A}^H(\mathbf{y}-\mathbf{A}\mathbf{X}_0)$
		\smallskip
		\State $\mathbb{K} \gets \mathrm{sort}(|\mathbf{Y}|)$, indices of $K$ largest $|\mathbf{Y}|$
		\State $\mathbf{X}_0\gets \mathbf{0}$, $\mathbf{X}_0\gets \mathbf{Y}$ for $k \in \mathbb{K}$, Hard Thresholding
		\EndFor
		
		\Output
		Reconstructed $\mathbf{X}_R=\mathbf{X}_0$, the set of positions $\mathbb{K}.$ 
	\end{algorithmic}
\end{algorithm}

\begin{figure*}[thb]
	\centering
		\includegraphics{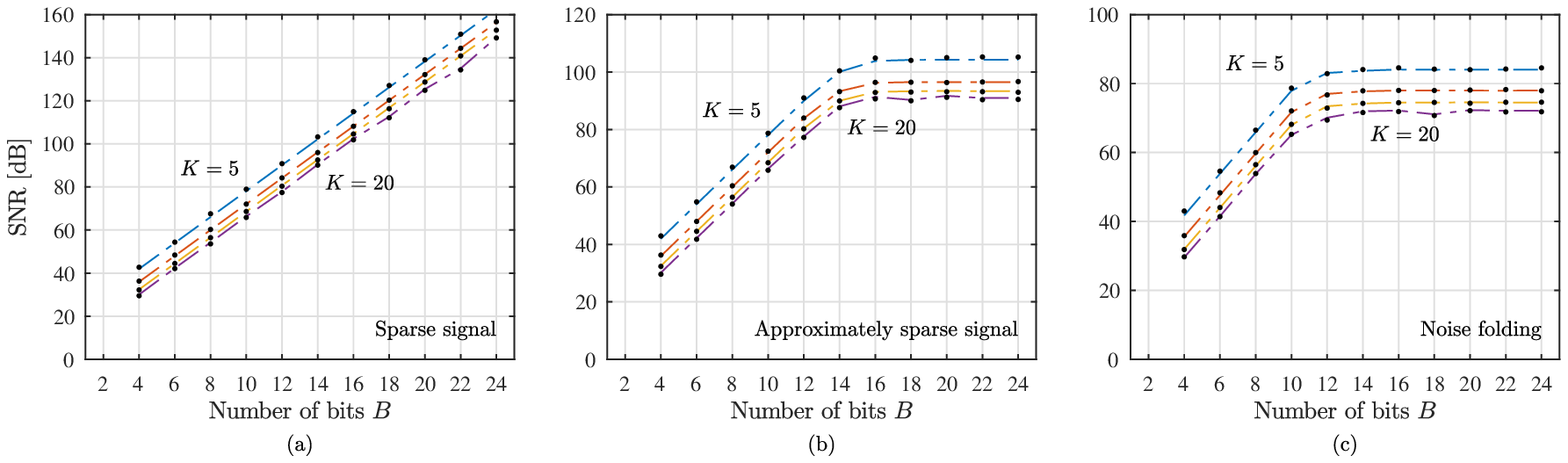}
	\caption{Reconstruction error for the partial Gaussian measurement matrix ($M=128$, $N=256$) when the Bayesian-based reconstruction algorithm is used. (a) Sparse signal with measurements quantized to fit the registers with $B$ bits for various sparsities. (b) Nonsparse signal with measurements quantized to fit the registers with $B$ bits for various sparsities. (c)  Nonsparse signals when both the measurements $\mathbf{y}$ and input coefficients $\mathbf{X}$ are quantized to $B$ bit fixed point registers (quantization noise folding).}	
	\label{Dig_Bayesian_gaus}
\end{figure*}

\medskip 
\textbf{Example 6:} In this example, the reconstruction of sparse and nonsparse signals, including noise folding, is performed using the Gaussian measurement matrix and the Bayesian-based method, \cite{BCS1,BCS2}, summarized in Algorithm \ref{BayesAlg}. The theoretical and statistical errors for the signals from Example 1 are shown in Fig. \ref{Dig_Bayesian_gaus}(a)-(c). Again, a high agreement between the statistical and theoretical results is obtained.

\begin{algorithm}[tb]
        \caption{Bayesian-based reconstruction}
        \label{BayesAlg}
        \begin{algorithmic}[1]
                \Input
                Vector $\mathbf{y}$, Matrix $\mathbf{A}$ 
                \Statex
                \State $\alpha_i \gets 1$
                        \Comment{For $i=1,2,\ldots,N$}
                \State $\sigma^2 \gets 1$               
                        \Comment{Initial estimate}
                \State $T_h=10^2$
                  \Comment{Threshold}
                \State $\mathbf{p}=[1,2,\ldots, N]^T$
                \Repeat
                \State $\mathbf{D} \gets $ diagonal matrix with $d_i$ values
                \State $\mathbf{\Sigma} \gets (\mathbf{A}^T \mathbf{A}/\sigma^2 +\mathbf{D})^{-1}$ 
                 \State $\mathbf{V} \gets \mathbf{\Sigma} \mathbf{A}^T \mathbf{y}/\sigma^2$
                 \State ${\gamma_i} \gets 1-d_i \Sigma_{ii}$
                     \Comment{For each $i$}
                 \State ${d_i} \gets \gamma_i / V_i$
                     \Comment{For each $i$}
                 \State $\displaystyle \sigma^2 \gets {\frac{\left\| \mathbf{y}-\mathbf{A}\mathbf{V} \right\|^2}{M-\sum_i \gamma_i}}$
                 \State $\mathbb{R} \gets\{ i : |d_i|>T_h \} $
                 \State Remove columns from matrix $\mathbf{A}$ selected by $\mathbb{R}$ 
                 \State Remove elements from array $d_i$ selected by $\mathbb{R}$ 
                 \State Remove elements from vector $\mathbf{p}$ selected by $\mathbb{R}$ 

                 \Until{stopping criterion is satisfied}
                 \State Reconstructed vector $\mathbf{X}$ nonzero coefficients are in vector $\mathbf{V}$ with corresponding positions in vector $\mathbf{p}$, 
                 $ X_{p_i}=V_i $
                
                \Statex
                \Output
                \Statex
                \begin{itemize}
                        \item Reconstructed signal vector $\mathbf{X}_R=\mathbf{V}$, the set of positions $\mathbb{K}=\mathbf{p}.$ 
                \end{itemize}
        \end{algorithmic}
\end{algorithm}

\section{Conclusions}
\label{Conc}

The effect of quantization noise to the reconstruction of signals under sparsity constraint is analyzed in this paper. If the measurements are not quantized, the reconstruction would be ideal and the error will be negligible. However, the quantization is a very useful step since the hardware realization of systems cannot store the exact values of samples. We derived the exact error of the reconstruction due to quantization noise. The cases when a signal is not strictly sparse is analyzed, as well as the noise folding effect. The reconstruction performance is validated on numerical examples, and compared to the statistical error calculation concluding a high agreement between them.

\appendix

\subsection*{Influence of quantization to the reconstruction condition}

Quantization noise can be included in the coherence index based relation for the reconstruction. The worst case amplitude of the considered normalized coefficient in the initial estimate is $ 1-(K-1)\mu - \nu \Delta/2 $ where 
$$\nu=\max_{k} \sum_{m=1}^{M}|a_m(k)| \le \sqrt{M}.$$
This inequality follows from the relation between the norm-two and the norm-one of a vector. 
For the partial DFT matrix, the random partial DFT matrix, and the Bernoulli matrix the equality $\nu=\sqrt{M}$ holds.
Following the same reasoning as in Remark \ref{SolUR}, we may conclude that at a position where the the ordinal coefficient $X(k)$ is zero-valued, 
in the worst case, the maximum possible disturbance is 
$K \mu + \sqrt{M} \Delta/2$. The detection of the strongest coefficient position is always successful if  $ 1-(K-1)\mu - \sqrt{M} \Delta/2 > K \mu + \sqrt{M} \Delta/2 $,
producing the condition for reconstruction 
\begin{equation}
K<\frac{1}{2}\left( 1+\frac{1}{\mu} -\frac{\sqrt{M} \Delta}{\mu}\right). \nonumber 
\end{equation} 
Influence of the quantization error to the uniqueness condition will be negligible if $\sqrt{M} \Delta \ll 1$ holds.

Coherence index based values guarantee exact reconstruction, however, they are pessimistic. More practical relations can be obtained by considering the  probabilistic analysis \cite{Srdjan_CS, LS_MB_audio}. 
The resulting disturbance in the initial estimate at the position $k \in \mathbb{K}$ due to the other coefficients and quantization noise behaves as a Gaussian random variable $\mathcal{N}(1,(K-1)\sigma_{\mu}^2+\sigma^2_\mathbf{e})$, for $K \gg 1$ . The initial estimate at $k \notin \mathbb{K}$  behaves as $\mathcal{N}(0,K\sigma_{\mu}^2+\sigma^2_\mathbf{e})$. Probabilistic analysis may provide approximative relations among $N$, $M$, and $K$, for a given probability. We have performed the statistical analysis with various measuremet matrices. The results of this analysis lead to the conclusion that for high probabilities of reconstruction we may neglect the quantization effect influence to the reconstruction condition for $B \ge 4$.  

Note that for large sparsities $K$ we have found that the reconstruction probability can be improved by increasing the upper limit for iterations in Algorithm \ref{Norm0Alg} for a few percents, with respect to the expected sparsity $K$. After the iterations are completed, the expected sparsity $K$ is used in the final reconstruction. This solves the problem that the iterative reconstruction in Algorithm \ref{Norm0Alg} cannot produce the exact result if it misses one of the nonzero coefficient positions during the iterative process for large $K$. 

%
%

%

\begin{IEEEbiography}[{\includegraphics[width=1in,height=1.25in,clip,keepaspectratio]{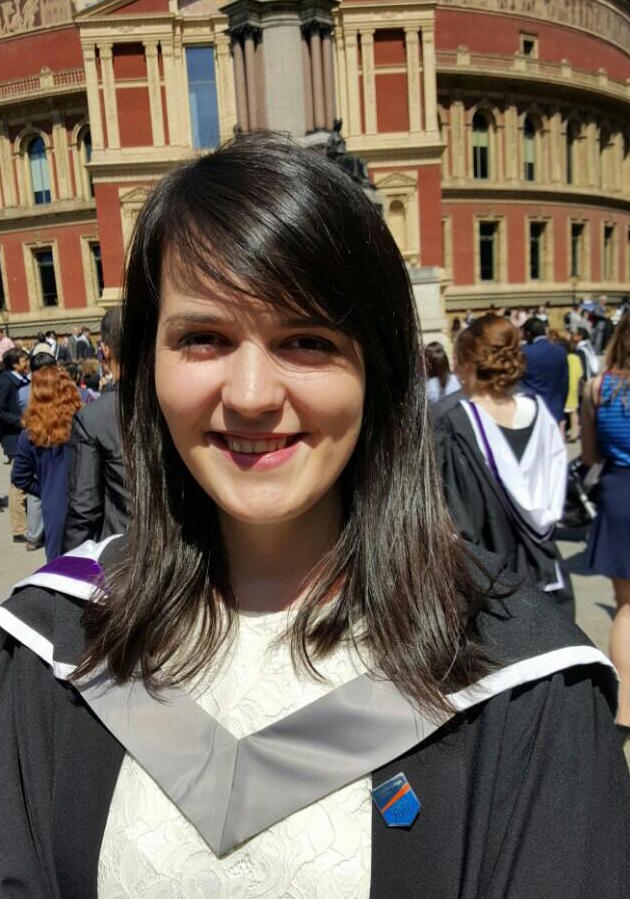}}]%
	{Isidora Stankovi\'{c}}
	was born in Podgorica, Montenegro in 1993. She received her BSc degree with First Class Honours in Electrical Engineering at the University of Westminster, London, United Kingdom, in 2014, and her MSc degree in Communications and Signal Processing at the Imperial College London, United Kingdom, in 2015. Currently, she is a PhD student on a double-degree program at the Faculty of Electrical Engineering, University of Montenegro and GIPSA Lab at the INP Grenoble, University of Grenoble Alpes. Her interest includes image processing, time-frequency analysis and compressive sensing algorithms. Isidora Stankovi\'c has published several papers in these areas. 
\end{IEEEbiography}

\begin{IEEEbiography}[{\includegraphics[width=1in,height=1.25in,clip,keepaspectratio]{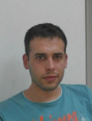}}]%
{Milo\v{s} Brajovi\'{c}} was born in Podgorica, Montenegro, in 1988. He received the B.S. and M.Sc. degrees in electrical engineering from the University of Montenegro, Podgorica, Montenegro, in 2011 and 2013, respectively. He is currently working toward the Ph.D. degree at the University of Montenegro. He is currently working as a teaching assistant at the University of Montenegro. He is a member of the Time-Frequency Signal Analysis Group, University of Montenegro, where he is involved in several research projects. His research interests include signal processing, time-frequency signal analysis, and compressive sensing. He has published several papers in these areas. 
\end{IEEEbiography}

\begin{IEEEbiography}[{\includegraphics[width=1in,height=1.25in,clip,keepaspectratio]{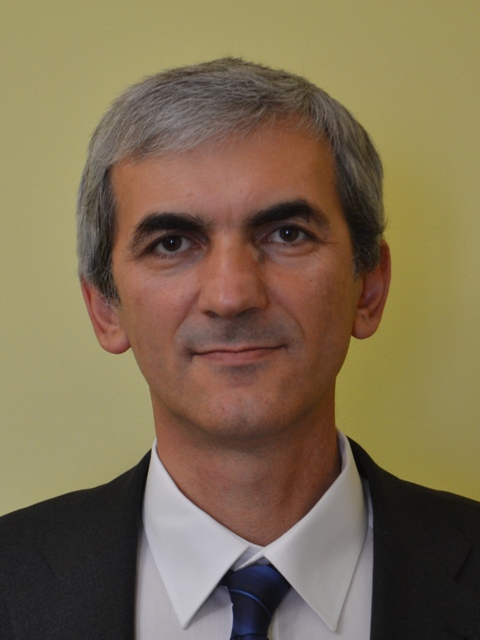}}]%
{Milo\v{s} Dakovi\'{c}} was born in 1970, Nik\v{s}i\'{c}, Montenegro.
He received the B.S. degree in 1996, the M.Sc. degree in 2001 and the Ph.D.
degree 2005, all at the University of Montenegro in EE. He is an associate
professor at the University of Montenegro. His research interests are signal
processing, time-frequency signal analysis and radar signal processing. He is
a member of the Time-Frequency Signal Analysis Group (www.tfsa.ac.me) at the
University of Montenegro where he was involved in several research projects
supported by Volkswagen foundation, Montenegrian Ministry of Science and
Canadian Government (DRDC).
\end{IEEEbiography}

\begin{IEEEbiography}[{\includegraphics[width=1in,height=1.25in,clip,keepaspectratio]{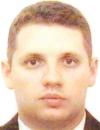}}]%
{Cornel Ioana}  received the Dipl.-Eng. degree in electrical engineering from the Romanian Military Technical Academy of Bucharest, Romania, in 1999 and the M.S. degree in telecommunication science and the Ph.D. degree in the electrical engineering field, both from University of Brest-France, in 2001 and 2003, respectively. Between 1999 and 2001, he activated as a Military Researcher in a research institute of the Romanian Ministry of Defense (METRA), Bucharest, Romania. Between 2003 and 2006, he worked as Researcher and Development Engineer in ENSIETA, Brest, France. Since 2006, he has been an Associate Professor-Researcher with the Grenoble Institute of Technology/GIPSA-lab. His current research activity deals with the signal processing methods adapted to the natural phenomena. His scientific interests are nonstationary signal processing, natural process characterization, underwater systems, electronic warfare, and real-time systems. 
\end{IEEEbiography}
\begin{IEEEbiography}[{\includegraphics[width=1in,height=1.25in,clip,keepaspectratio]{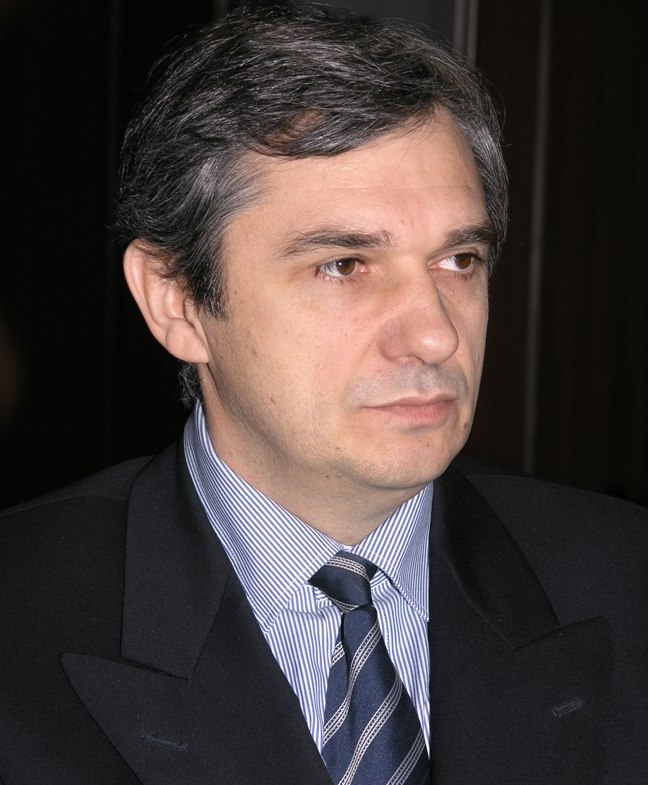}}]%
{Ljubi\v{s}a Stankovi\'{c}}
(\textit{M'91--SM'96--F'12}) was born in
Montenegro in 1960. He received the B.S. degree in EE from the University
of
Montenegro (UoM), the M.S. degree in Communications from the University of
Belgrade and the Ph.D. in Theory of Electromagnetic Waves from the UoM.\
As a
Fulbright grantee, he spent 1984-1985 academic year at the Worcester
Polytechnic Institute, USA. Since 1982, he has been on the faculty at the
UoM,
where he has been a full professor since 1995. In 1997-1999, he was on leave
at the Ruhr University Bochum, Germany, supported by the AvH Foundation.
At
the beginning of 2001, he was at the Technische Universiteit Eindhoven, The
Netherlands, as a visiting professor. He was vice-president of Montenegro
1989-90. During the period of 2003-2008, he was Rector of the UoM. He was
Ambassador of Montenegro to the UK, Ireland, and Iceland from 2010 to 2015.
His current interests
are in Signal Processing. He published about 400 technical papers, more than
160 of them in the leading journals, mainly the IEEE editions. Prof.
Stankovi\'{c} received the highest state award of Montenegro in 1997, for
scientific achievements. He was a member the IEEE SPS Technical Committee
on
Theory and Methods, an Associate Editor of the \textit{IEEE Transactions
on
Image Processing}, the \textit{IEEE Signal Processing Letters}, \textit{IEEE
Transactions on Signal Processing},
and numerous special issues of journals. Prof. Stankovi\'{c} is a member
of Editorial Board of \textit{Signal Processing} and a Senior area Editor of the \textit{IEEE Transactions
on
Image Processing}. 
He is a member of the 
National Academy of Science and Arts of Montenegro (CANU) since 1996 and its vice-president since 2016. He is 
a
member of Academia Europea (2012). Stankovic (with coauthors) won the Best paper award from the European Association for Signal Processing (EURASIP) in 2017 for a paper published in the \textit{Signal Processing} journal.
\end{IEEEbiography}

\end{document}